\documentclass{article}

\PassOptionsToPackage{numbers,sort&compress}{natbib}
\usepackage[final]{neurips_2023}
\usepackage{tablefootnote}

\usepackage[utf8]{inputenc} 
\usepackage[T1]{fontenc}    
\usepackage{hyperref}       
\usepackage{url}            
\usepackage{booktabs}       
\usepackage{amsfonts}       
\usepackage{nicefrac}       
\usepackage{microtype}      
\usepackage{xcolor}         
\usepackage{graphicx}

\usepackage{setspace}
\usepackage{algorithm}
\usepackage{algorithmic}
\newcommand\norm[1]{{\left\lVert#1\right\rVert}} 

\usepackage{multirow}
\usepackage{makecell}
\usepackage{amsmath,amssymb}
\DeclareMathAlphabet{\bbold}{U}{bbold}{m}{n}
\newcommand{\id}{\ensuremath{\bbold{1}}}

\title{Mixed Continuous and Categorical Flow Matching for 3D De Novo Molecule Generation}

\author{%
  Ian Dunn \\
  Dept. of Computational \& Systems Biology\\
  University of Pittsburgh\\
  Pittsburgh, PA 15260 \\
  \texttt{ian.dunn@pitt.edu} \\
  \And
  David Ryan Koes \\
  Dept. of Computational \& Systems Biology\\
  University of Pittsburgh\\
  Pittsburgh, PA 15260 \\
  \texttt{dkoes@pitt.edu} \\
}

\begin{document}

\maketitle

\begin{abstract}
    Deep generative models that produce novel molecular structures have the potential to facilitate chemical discovery. Diffusion models currently achieve state of the art performance for 3D molecule generation. In this work, we explore the use of flow matching, a recently proposed generative modeling framework that generalizes diffusion models, for the task of de novo molecule generation. Flow matching provides  flexibility in model design; however, the framework is predicated on the assumption of continuously-valued data. 3D de novo molecule generation requires jointly sampling continuous and categorical variables such as atom position and atom type. We extend the flow matching framework to categorical data by constructing flows that are constrained to exist on a continuous representation of categorical data known as the probability simplex. We call this extension SimplexFlow. We explore the use of SimplexFlow for de novo molecule generation. However, we find that, in practice, a simpler approach that makes no accommodations for the categorical nature of the data yields equivalent or superior performance. As a result of these experiments, we present FlowMol, a flow matching model for 3D de novo generative model that achieves improved performance over prior flow matching methods, and we raise important questions about the design of prior distributions for achieving strong performance in flow matching models. Code and trained models for reproducing this work are available at \url{https://github.com/dunni3/FlowMol}.
\end{abstract}

\begin{figure}[t]
    \centering
    \label{fig:ga}
    \includegraphics[scale=0.9]{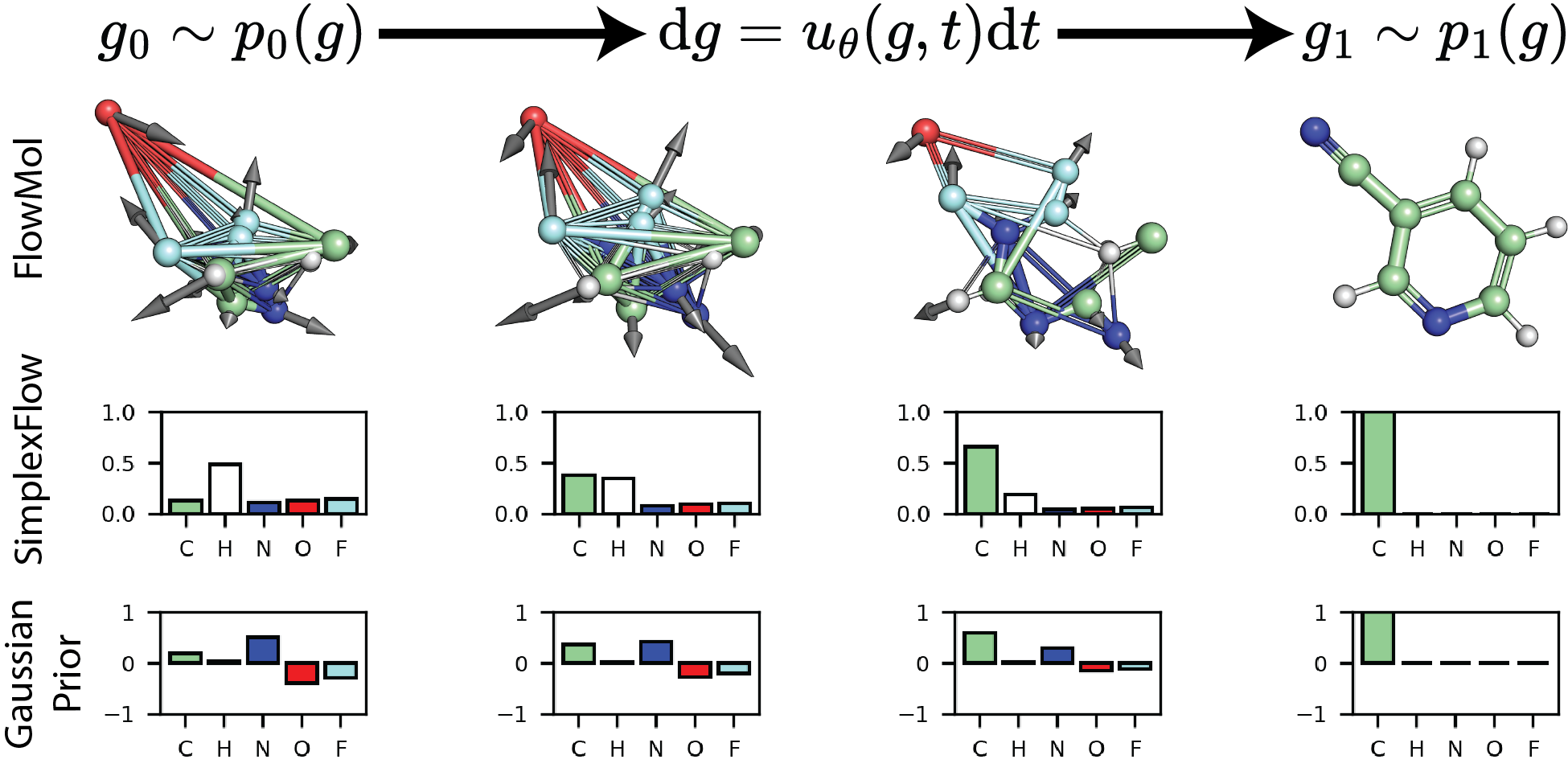}
    \caption{\textbf{Overview of FlowMol} \textit{Top:} We adapt the flow matching framework for unconditional 3D molecule generation. An ordinary differential equation parameterized by a graph neural network transforms a prior distribution over atom positions, types, charges, and bond orders to the distribution of valid molecules. Black arrows show the instantaneous direction of the ODE on atom positions. \textit{Middle:} Trajectory of the atom type vector for a single atom under SimplexFlow, a variant of flow matching developed for categorical variables. Atom type flows lie on the probability simplex. \textit{Bottom:} Trajectory of an atom type vector starting from a Gaussian prior. This approach does not respect the categorical nature of the data; however, we find it yields superior performance to SimplexFlow.}
\end{figure}

\section{Introduction}

Deep generative models that can directly sample molecular structures with desired properties have the potential to accelerate chemical discovery by reducing or eliminating the need to engage in resource-intensive screening-based based discovery paradigms. Moreover, generative models may also improve chemical discovery by enabling multi-objective design of chemical matter. In pursuit of this idea, there has been recent interest in developing generative models for the design of small-molecule therapeutics \cite{huang_dual_2024,guan_3d_2023,schneuing_structure-based_2023,peng_pocket2mol_2022,liu_generating_2022,torge_diffhopp_2023,igashov_equivariant_2024,dunn_accelerating_2023}, proteins \cite{watson_novo_2023,bennett_atomically_2024,ingraham_illuminating_2023}, and materials \cite{zeni_mattergen_2024}. State of the art performance in these tasks is presently achieved by applying diffusion models \cite{sohl-dickstein_deep_2015,ho_denoising_2020,song_score-based_2021} to point cloud representations of molecular structures. 

Flow matching, a recently proposed generative modeling framework \cite{lipman_flow_2023, tong_improving_2023, albergo_stochastic_2023, liu_flow_2022}, generalizes diffusion models. Under diffusion models, the transformation of prior samples to data is formulated as a reversal of a predefined forward process. The forward process is a Markov chain or differential equation that must converge to a tractable stationary distribution as $t \to \infty$; this requirement constrains the viable options for forward/reverse processes and prior distributions. In contrast, flow matching prescribes a method for directly learning a differential equation that maps samples from nearly arbitrary distributions. In doing so, flow matching permits valuable flexibility when designing models for specific applications. For example, \citet{jing_alphafold_2024} and \citet{stark_harmonic_2024} make use of the fact that flow matching allows arbitrary prior distributions to design models whose priors are closer to realistic 3D molecular conformations than a Gaussian prior.

In this work we explore the application of flow matching to 3D de novo small molecule generation. We adapt the approach of state of the art diffusion models for this task \cite{huang_learning_2023,vignac_midi_2023, peng_moldiff_2023} to the flow matching framework. This approach entails predicting atom positions, atom types (chemical elements), formal charges, and bond orders between all pairs of atoms. All of these variables are categorical with the exception of atom positions. Therefore, molecule generation requires sampling from a joint distribution of continuous and categorical variables. 

Effectively adapting flow matching for this mixed continuous/categorical generative task may be non-trivial because the flow matching framework is predicated on the assumption of continuously valued data. In this work, we extend the flow matching framework to categorical data by constructing flows that are constrained to exist on a continuous representation of categorical data known as the probability simplex. We call this extension SimplexFlow. We present a model for de novo small-molecule generation that uses SimplexFlows to generate categorical features.

This work was motivated by the intuition that designing a generative process that respects the categorical nature of the data it operates on may yield improved performance; however, our empirical results contradict this intuition. We show that in practice, a simpler approach that makes no accommodations for the categorical nature of the data yields superior performance to a de novo model using SimplexFlow. Our final flow matching model for molecule generation, FlowMol, achieves improved performance over existing flow matching methods for molecule generation and is competitive with state of the art diffusion models while exhibiting a >10-fold reduction in inference time. 

\section{Background}

\subsection{Discrete Diffusion}

The original formulation of diffusion models \cite{sohl-dickstein_deep_2015} was defined in terms of a Markov chain of random variables that converged to a tractable stationary distribution in the limit of an infinite number of steps in the Markov chain. This formulation made no assumptions about the sample space of the random variables modeled, allowing for natural extensions to discrete data \cite{hoogeboom_argmax_2021, campbell_continuous_2022, austin_structured_2023}. 

A separate formulation of diffusion models as continuous-time stochastic differential equations (SDE)\cite{song_score-based_2021} became popular in the literature. The SDE formulation of diffusion models is dependent on the assumption of having continuously-valued data. Similar to our approach, there is a line of work developing SDE-based diffusion models that operate on continuous representations of discrete data. Several works developed diffusion models where diffusion trajectories were constrained to the simplex \cite{richemond_categorical_2022, floto_diffusion_2023, avdeyev_dirichlet_2023}. An alternative approach is to embed categorical features into a continuous latent space and train diffusion models on the embeddings \cite{dieleman_continuous_2022}.

\subsection{De Novo Molecule Generation} \label{sec: de novo background}

Initial attempts at de novo molecule generation focused on generating either textual representations (SMILES strings) \cite{grisoni_bidirectional_2020,gomez-bombarelli_automatic_2018,dai_syntax-directed_2018} or 2D molecular graphs \cite{jin_junction_2019,liu_constrained_2019,shi_graphaf_2020,you_graph_2019}: molecular representations that exclude all information about 3D structure. Subsequent approaches were developed for 3D molecule generation using a variety of molecular representations and generative paradigms \cite{ragoza_learning_2020,ragoza_generating_2022,gebauer_symmetry-adapted_2020,luo_autoregressive_2021,satorras_en_2022}.

\citet{hoogeboom_equivariant_2022} proposed the first diffusion model for 3D molecule generation, which yielded superior performance over previous approaches. Molecules are represented in \citet{hoogeboom_equivariant_2022} by attributed point clouds where each atom has a position in space and type. A continuous diffusion process is defined for both atom positions and types where the prior for both is a standard Gaussian distribution. A purported weakness of this approach is that atom connectivity is not predicted by the model and must be inferred in a post-processing step. Several concurrent works sought to address these issues by predicting bond order in addition to atom positions/types: \citet{vignac_midi_2023, huang_learning_2023, peng_moldiff_2023, hua_mudiff_2024}. These models report substantially improved performance over \citet{hoogeboom_equivariant_2022}. Three of these four concurrent works (\citet{vignac_midi_2023,peng_moldiff_2023,hua_mudiff_2024}) use discrete diffusion processes for categorical features and attribute (in part) their improved model performance to the use of discrete diffusion.; however, only \citet{peng_moldiff_2023} presents an ablation study isolating the effect of discrete diffusion. Moreover, \citet{huang_learning_2023} uses only continuous diffusion processes and reports superior performance. This suggests that while predicting graph connectivity provides performance benefits, the utility of discrete diffusion for molecule generation is less clear. \citet{vignac_midi_2023} and \citet{huang_learning_2023} fully specify the molecular structure by also predicting atom formal charges and the presence of hydrogen atoms; for this reason, these works are  the most similar to the model presented here.

\subsection{Flow-Matching for De Novo Molecule Generation} \label{sec:fmdenovo}

To our knowledge, \citet{song_equivariant_2023} is the only existing work that performs de novo molecule generation with flow matching. Molecules are represented as point clouds where each atom has a position in space and an atom type. The final molecule structure is inferred after the inference procedure. The prior distribution for atom type vectors is a standard Gaussian distribution, and so the generative process does not have any inductive biases to respect the discrete nature of the data. This work can be viewed as the flow matching analog of \citet{hoogeboom_equivariant_2022}.

\subsection{Flow Matching for Discrete Data}

Concurrent work \cite{stark_dirichlet_2024} developed a variant of flow matching on the simplex which we refer to as Dirichlet Flows. In Dirichlet Flows, conditional probability paths are only conditioned on $x_1$ and, as a result, do not permit arbitrary choices of the prior and must use a uniform distribution over the simplex. In contrast, our formulation permits the use of any prior distribution. \citet{stark_dirichlet_2024} identify problems with the choice of commonly used conditional vector fields that limit performance on variables with a large number of categories. They propose an alternative choice of conditional probability paths that alleviate this issue.

There are also other works which develop flow matching variants for discrete data. \citet{boll_generative_2024} equip the simplex with the Fisher-Rao metric to form a Riemannian manifold, and apply Riemannian Flow Matching \cite{chen_flow_2024} to this manifold. \citet{campbell_generative_2024} develop a flow matching method for discrete data built on continuous-time Markov chains.

Importantly, none of the aforementioned works, which present methods for training flow matching models for categorical data, benchmark their model performance against simpler flow matching models that do not account for the categorical nature of their data.

\subsection{Flow Matching} \label{sec:fm-background}

Flow matching \cite{liu_flow_2022, tong_improving_2023, albergo_stochastic_2023, lipman_flow_2023} is a new generative modeling framework that generalizes diffusion models. Flow matching permits useful design flexibility in the choice of prior of and nature of the map between two distributions. Flow matching is also conceptually simpler than diffusion and permits substantially faster inference. We briefly describe the flow matching framework here. 

An ordinary differential equation (ODE) that exists on $\mathbb{R}^d$ is defined by a smooth, time-dependent vector-field $u(x,t): \mathbb{R}^d \times [0,1] \to \mathbb{R}^d$.

\begin{equation} \label{eq:fm-ode}
    \frac{dx}{dt} = u(x,t)
\end{equation}

Note that we only consider this ODE on the time interval $[0,1]$. For simplicity we will use $u_t(x)$ interchangeably with $u(x,t)$. Given a probability distribution over initial positions $x_0 \sim p_0(x)$, the ODE \eqref{eq:fm-ode} induces time dependent probability distributions $p_t(x)$. The objective in flow matching is to approximate a vector field $u_t(x)$ that pushes a source distribution $p_0(x)$ to a desired target distribution $p_1(x)$. A neural network $u_\theta$ can be regressed to the vector field $u_t$ by minimizing the Flow Matching loss.

\begin{equation} \label{fm-loss}
    \mathcal{L}_{FM} = \mathbb{E}_{x \sim p_t} \norm{ u_\theta(x,t) - u_t(x) }^2
\end{equation}

Computing $\mathcal{L}_{FM}$ requires access to $u_t$ and $p_t$, quantities that are typically intractable. Flow matching provides a method for approximating $u_t(x)$ without  having access to it. If we consider the probability path $p_t(x)$ to be a mixture of \textit{conditional} probability paths $p_t(x|z)$:

\begin{equation}
    p_t(x) = \int{ p_t(x|z)p(z)dz }
\end{equation}

and we know the form of the the conditional vector fields $u_t(x|z)$ that produce $p_t(x|z)$, then the marginal vector field $u_t(x)$ can be defined as a mixture of conditional vector fields:

\begin{equation}
    u_t(x) = \mathbb{E}_{p(z)} \frac{u_t(x|z)p_t(x|z)}{p_t(x)}
\end{equation}

We still cannot compute $u_t(x)$ but the neural network $u_\theta$ that is the minimizer of $\mathcal{L}_{FM}$ is also the minimizer of the Conditional Flow Matching (CFM) loss defined in \eqref{eq:cfm-loss}

\begin{equation} \label{eq:cfm-loss}
    \mathcal{L}_{CFM} = \mathbb{E}_{p(z), p_t(x|z), t\sim \mathcal{U}(0,1)} \norm{ u_\theta(x,t) - u_t(x|z) }^2
\end{equation}

That is, regressing to conditional vector fields, in expectation, is equivalent to regressing to the marginal vector field. The remaining design choices for a flow matching model are the choice of conditioning variable $z$, conditional probability paths $p_t(x|z)$, and conditional vector fields $u_t(x|z)$.

\section{Methods}

\subsection{Problem Setting} \label{sec:problem setting}


We represent a molecule with $N$ atoms as a fully-connected graph. Each atom is a node in the graph. Every atom has a position in space $X = \{x_i\}_{i=1}^N \in \mathbb{R}^{N \times 3}$, an atom type (in this case the atomic element) $A = \{a_i\}_{i=1}^N \in \mathbb{R}^{N \times n_a} $, and a formal charge $C = \{c_i\}_{i=1}^N \in \mathbb{R}^{N \times n_c}$. Additionally, every pair of atoms has a bond order $E = \{  e_{ij} \forall i,j \in [N] | i \neq j \} \in \mathbb{R}^{(N^2 - N) \times n_e} $. Where $n_a, n_c, n_e$ are the number of possible atom types, charges, and bond orders; these are categorical variables represented by one-hot vectors.  For brevity, we denote a molecule by the symbol $g$, which can be thought of as a tuple of the constituent data types $g = (X, A, C, E)$. 

There is no closed-form expression or analytical technique for sampling the distribution of realistic molecules $p(g)$. We seek to train a flow matching model to sample this distribution. Concretely, we choose the the target distribution that is the distribution of valid 3D molecules $p_1(g) = p(g)$. Our choice of prior $p_0(g)$ is described in Section \ref{sec:priors}.

Our strategy for adapting flow matching to molecular structure is one that mimics prior work on applying diffusion and flow-based generative models to molecular structure. That is, we define conditional vector fields and conditional probability paths for each data modality and jointly regress one neural network for all data modalities. Our total loss is a weighted combination of CFM losses from \eqref{eq:cfm-loss}:

\begin{equation}
    \mathcal{L} = \eta_X \mathcal{L}_{X} + \eta_A \mathcal{L}_{A} + \eta_C \mathcal{L}_{C} + \eta_E \mathcal{L}_{E}
\end{equation}

Where $(\eta_{X}, \eta_{A}, \eta_{C}, \eta_{E})$ are scalars weighting the relative contribution of each loss term. We set these values to $(3, 0.4, 1, 2)$ as was done in \citet{vignac_midi_2023}. Our specific choice of conditional vector fields and probability paths is described in Section \ref{sec:our-fm}. In practice, we use a variant of the CFM objective called the endpoint-parameterized objective that we present in Section \ref{sec:ep}. These choices are used to in turn to design SimplexFlow, our method of performing flow matching for categorical variables, which is described in Section \ref{sec:simpflow}.

\subsection{Flow Matching with Temporally Non-Linear Interpolants} \label{sec:our-fm}

We choose the conditioning variable to be the initial and final states of a trajectory: $z = (g_0, g_1)$. We choose the conditional probability path to be a Dirac density placed on a ``straight'' line connecting these states $p_t(g|g_0,g_1) = \delta(g - (1 - \alpha_t)g_0 - \alpha_t g_1)$. This particular choice of conditional vector fields and probability paths gives us the freedom to choose any prior distribution $p_0(g)$ \cite{tong_improving_2023, albergo_stochastic_2023, liu_constrained_2019}. Our choice of $p_t(g|g_0,g_1)$ is equivalent to defining a deterministic interpolant:

\begin{equation} \label{eq:flex-interpolant}
    g_t = (1 - \alpha_t)g_0 + \alpha_t g_1
\end{equation}

where $\alpha_t: [0,1] \to [0,1]$ is a function that takes $t$ as input and returns a value between 0 and 1. The rate at which  a molecule from the prior distribution $g_0$ is transformed into a valid molecule $g_1$ can be controlled by choice of $\alpha_t$, which we name the ``interpolant schedule.''\footnote{This is intended to be analogous to noise schedules for diffusion models.} We define separate interpolant schedules for each data type comprising a molecule: $\alpha_t = (\alpha_t^X, \alpha_t^A, \alpha_t^C, \alpha_t^E)$. Taking inspiration from \citet{vignac_midi_2023}, we define a cosine interpolant schedule:

\begin{equation} \label{eq:cos-sched}
    \alpha_t = 1 - \cos^2 \left(\frac{\pi}{2}t^{\nu}
    \right)
\end{equation}

where different values of $\nu$ are set for atom positions, types, charges, and bond orders. The interpolant \eqref{eq:flex-interpolant} gives rise to conditional vector fields of the form:

\begin{equation} \label{eq:flex-vf-simple}
    u(g_t|g_0,g_1) = \alpha'_t(g_1 - g_0)
\end{equation}

Where $\alpha'_t$ is the time derivative of $\alpha_t$.

\subsection{Endpoint Parameterization} \label{sec:ep}

By solving  \eqref{eq:flex-interpolant} for $g_0$ and substituting this expression into \eqref{eq:flex-vf-simple} we obtain an alternate form of the conditional vector field. 

\begin{equation} \label{eq:ep-cond-vec-field}
    u(g_t|g_0,g_1) = \frac{\alpha'_t}{1-\alpha_t}(g_1 - g_t)
\end{equation}

As described in Section \ref{sec:fm-background}, the typical flow matching procedure is to regress a neural network $u_\theta(g_t)$ directly to conditional vector fields by minimizing the CFM loss \eqref{eq:cfm-loss}. Instead, we apply a reparameterization initially proposed by \citet{jing_alphafold_2024}:

\begin{equation} \label{eq:vec-field-reparam}
    u_\theta(g_t) = \frac{\alpha'_t}{1-\alpha_t} \left( \hat{g}_1(g_t) - g_t \right) 
\end{equation}

By substituting \eqref{eq:vec-field-reparam} and \eqref{eq:ep-cond-vec-field} into \eqref{eq:cfm-loss}, we obtain our endpoint-parameterized objective

\begin{equation} \label{eq:endpoint-objective}
    \mathcal{L}_{EP} = \mathbb{E}_{t, g_t} \left[  \frac{\alpha'_t}{1 - \alpha_t}  || \hat{g}_1(g_t) - g_1  || \right]
\end{equation}

Therefore our objective becomes to train a neural network that predicts valid molecular structures given samples from a conditional probability path $\hat{g}_1(g_t)$. This is particularly advantageous when operating on categorical data, as placing a softmax layer on model outputs constrains the domain of model outputs to the simplex. Empirically, we find that the endpoint objective yields better performance than the vector field regression objective \eqref{eq:cfm-loss} for the task of molecule generation. Moreover, we leverage the theoretical guarantee that our predicted endpoint for categorical data lie on the simplex to ensure our flows lie on the simplex.

In practice, the interpolant-dependent loss weight $\frac{\alpha'_t}{1 - \alpha_t}$ produces unreasonably large values as $\alpha_t \to 1$. We replace this term with a time-dependent loss function inspired by \citet{le_navigating_2023}: $w(t) = \min(\max(0.005, \frac{\alpha_t}{1-\alpha_t}), 1.5)$. For categorical variables we use a cross entropy loss rather than the L2 norm shown in \eqref{eq:endpoint-objective}.

\subsection{SimplexFlow} \label{sec:simpflow}

To design flow matching for categorical data, our strategy is to define a continuous representation of categorical variables, and then construct a flow matching model where flows are constrained to this representation. We choose the d-dimensional probability simplex $\mathcal{S}^d$ as the continuous representation of a $d$-categorical variable.

\begin{equation}
    \mathcal{S}^d = \left\{ x \in \mathbb{R}^d | x_i > 0, \id \cdot x=1  \right\}
\end{equation}

A $d$-categorical variable $x_1 \in \{ 1,2,\dots,d \}$ can be converted to a point on $\mathcal{S}^d$ via one-hot encoding. Correspondingly, the categorical distribution $p_1(x) = \mathcal{C}(q)$ can be converted to a distribution on $\mathcal{S}^d$ as:

\begin{equation} \label{eq:simplex-data-dist}
    p_1(x) = \sum_{i=1}^d q_i\delta(x - e_i)
\end{equation}

where $e_i$ is the $i^{th}$ vertex of the simplex and $q_i$ is the probability of $x$ belonging to the $i^{th}$ category. If we choose a prior distribution $p_0(x)$ such that $\mathrm{supp}(p_0) = \mathcal{S}^d$, then all conditional probability paths produced by the interpolant \eqref{eq:flex-interpolant} will lie on the simplex. This is because the simplex is closed under linear interpolation (see Appendix \ref{ap:closed simplex}) and the conditional trajectories are obtained by linearly interpolating between two points on the simplex ($x_0,x_1 \in \mathcal{S}^d$).

Although choosing $(p_0, p_1)$ with support on the simplex results in conditional trajectories on the simplex, training a flow under the vector field objective \eqref{eq:cfm-loss} provides no guarantee that trajectories produced by the learned vector field lie on the simplex. However, training a flow matching model under the endpoint parameterization (Section \ref{sec:ep}) enables us to guarantee by construction that generated flows lie on the simplex; proof of this is provided in Appendix \ref{ap:simplex-proof}.

\subsection{Priors} \label{sec:priors}

We define the prior distribution for a molecule as a composition of independent samples for each atom and pair of atoms. Our prior distributions take the form:

\begin{equation} \label{eq:prior-form}
    p_0(g) = p_0(x, a, c, e) = \prod_{i=1}^{N_{\mathrm{atoms}}} p_0(x_i)p_0(a_i)p_0(c_i) \times \prod_{i,j<i}^{N_{\mathrm{atoms}}} p_0(e_{ij})
\end{equation}

Our choice of conditional trajectory \eqref{eq:flex-interpolant} permits the choice of any prior distribution. SimplexFlow places the constraint that the prior distribution for categorical variables have support bounded to the simplex. 

We always set $p_0(x_i) = \mathcal{N}(x_i|0,\mathbb{I})$; atom positions are independently sampled from a standard Gaussian distribution. We explore the use of several prior distributions for categorical variables $a_i,c_i,e_{ij}$. We experiment with three different categorical priors for SimplexFlow. The \textbf{uniform-simplex} prior is a uniform distribution over the simplex; the simplest choice for a categorical prior. This choice is analogous to the ``Linear FM'' model described in \cite{stark_dirichlet_2024}. The \textbf{marginal-simplex} prior is designed to be ``closer'' to the data distribution by using marginal distributions observed in the training data. Specifically, we replace $p_0(a_i)p_0(c_i)$ and $p_0(e_{ij})$ in \eqref{eq:prior-form} with $p_1(a_i,c_i)$ and $p_1(e_{ij})$, respectively. Finally, for the \textbf{barycenter} prior, categorical variables are placed at the barycenter of the simplex; the point in the center of the simplex assigning equal probability to all categories. The intuition behind the barycenter prior is all categorical variables will be ``undecided'' at $t=0$. 




In practice, the model fails when the prior distributions for categorical variables only have density on a small, fixed number of points on the simplex; this is the case for the marginal-simplex and barycenter priors. We find that "blurring" the prior samples for categorical variables significantly improves performance. That is, Gaussian noise is added to the samples before they are projected back onto the simplex. 


\subsection{Optimal Transport Alignment}

Previous work \cite{tong_improving_2023} has shown that aligning prior and target samples via optimal transport significantly improves the performance of flow matching by minimizing the extent to which conditional trajectories intersect. When performing flow matching on molecular structure, this consists of computing the optimal permutation of node ordering and the rigid-body alignment of atom positions \cite{klein_equivariant_2023,song_equivariant_2023}. We apply the same alignment between target and prior positions at training time. This also ensures that prior positions $p_0(X)$ and target positions $p_1(X)$ effectively exist in the center of mass free subspace proposed in \citet{hoogeboom_equivariant_2022} that renders the target density $p_1(g)$ invariant to translations.

\subsection{Model Architecture}

Molecules are treated as fully-connected graphs. The model is designed to accept a sample $g_t$ and predict the final destination molecule $g_1$. Within the neural network, molecular features are grouped into node positions, node scalar features, node vector features, and edge features. Node positions are identical to atom positions discussed in Section \ref{sec:problem setting}. Node scalar features are a concatenation of atom type and atom charge. Node vector features are geometric vectors (vectors with rotation order 1) that are relative to the node position. Node vector features are initialized to zero vectors. Molecular features are iteratively updated by passing $g_t$ through several Molecule Update Blocks. A Molecule Update Block uses Geometric Vector Perceptrons (GVPs) \cite{jing_equivariant_2021} to handle vector features. Molecule Update Blocks are composed of three components: a node feature update (NFU), node position update (NPU) and edge feature update (EFU). The NFU uses a message-passing graph convolution to update node features. The NPU and EFU blocks are node and edge-wise operations, respectively. Following several molecule update blocks, predictions of the final categorical features ($\hat{A}_1, \hat{C}_1, 
\hat{E}_1$) are generated by passing node and edge features through shallow node-wise and edge-wise multi layer perceptrons (MLPs). For models using endpoint parameterization, these MLPs include softmax activations. The model architecture is visualized in Figure \ref{fig:arch} and explained in detail in Appendix \ref{ap:arch}.

\begin{figure}[t]
    \centering
    \includegraphics[scale=0.99]{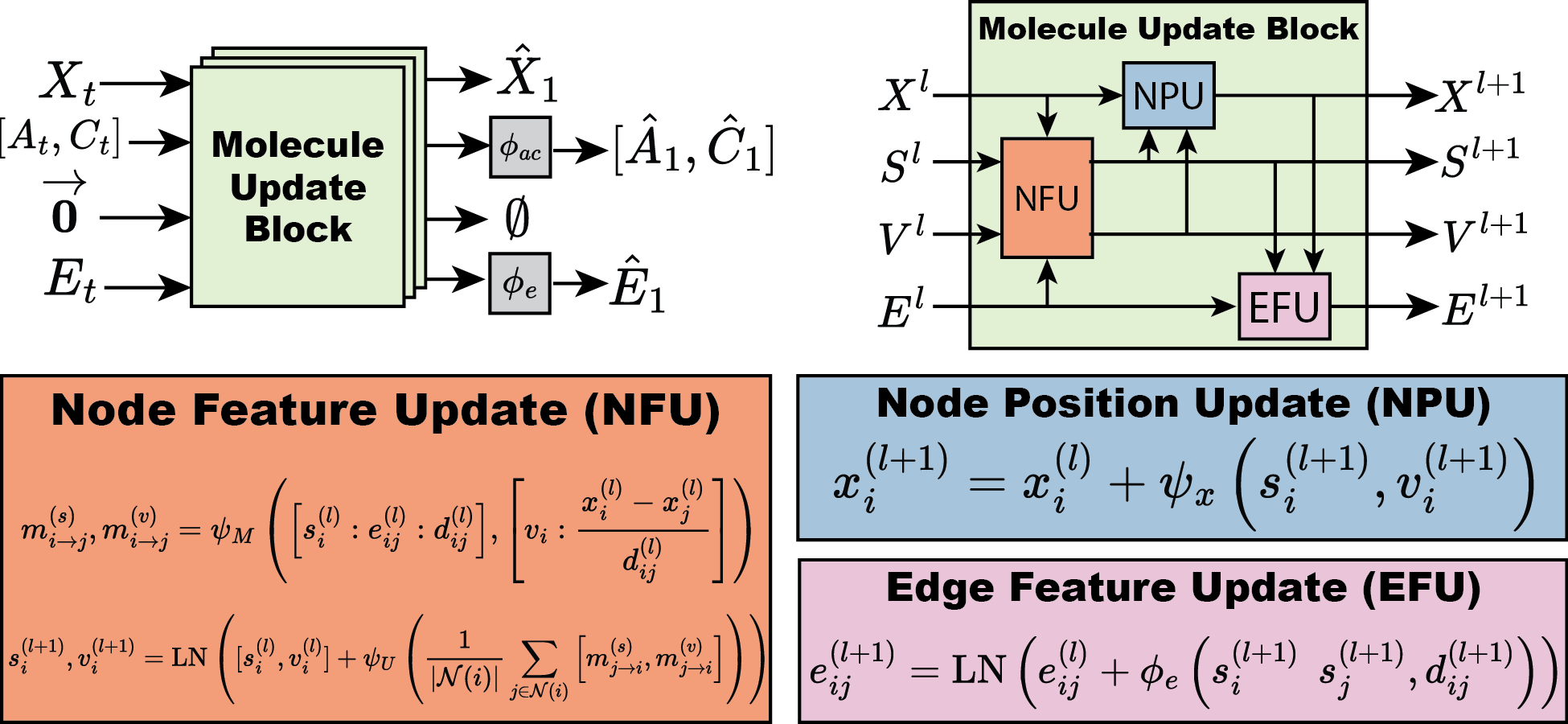}
    \caption{\textbf{FlowMol Architecture} \textit{Top left}: An input molecular graph $g_t$ is transformed into a predicted final molecular graph $g_1$ by being passed through multiple molelcule update blocks. \textit{Top right:} A molecule update block uses NFU, NPU, and EFU sub-components to update all molecular features. \textit{Bottom:} Update equations for graph features. $\phi$ and $\psi$ is used to denote MLPs and GVPs, respectively.  }
    \label{fig:arch}
\end{figure}

In practice, graphs are directed. For every pair of atoms $i,j$ there exists edges in both directions: $i \rightarrow j$ and $j \rightarrow i$. When predicting the final bond orders $\hat{E}_1$ for an edge, we ensure that one prediction is made per pair of atoms and that this prediction is invariant to permutations of the atom indexing. This is accomplished by making our prediction from the sum of the learned bond features. That is, $\hat{e}_1^{ij} = MLP( e_{ij} + e_{ji} )$. 

GVPs, as they were originally designed, predict vector quantities that are E(3)-equivariant. We introduce a variant of GVP that is made SE(3)-equivariant by the addition of cross product operations. The cross product is equivariant to rotations and translations of input vectors but not reflections. As a result, the learned density $p_1(g)$ is invariant to rotations and translations but not reflections.
In other words, FlowMol is sensitive to chirality. Empirically we find that the addition of cross product operations to GVP improves performance. \citet{schneuing_structure-based_2023} proposed the addition of a cross product operation to the EGNN architecture \cite{satorras_en_2022-1}; we adopt this idea for GVP. We refer the reader to Appendix F of \citet{schneuing_structure-based_2023} for a detailed discussion of the equivariance of cross products. Our cross product variant of GVP is described in Appendix \ref{sec:gvp}.

\section{Experiments}

\subsection{Datasets}

We train on QM9 \cite{ruddigkeit_enumeration_2012,ramakrishnan_quantum_2014} and GEOM-Drugs \cite{axelrod_geom_2022} using explicit hydrogens. QM9 contains 124k small molecules, each with one 3D conformation. GEOM-Drugs contains approximately 300k larger, drug-like molecules with multiple conformers for each molecule. Molecules in QM9 have an average of 18 atoms a max of 29 while those in GEOM-Drugs have an average of 44 atoms and a max of 181. We use the same dataset splits as \citet{vignac_midi_2023}. 

We chose to use explicit hydrogens because it is a more difficult learning task. By predicting explicit hydrogens in combination with atom types, bond orders, and formal charges, there is a 1-to-1 mapping from model outputs to molecules. If any one of these components were removed from the generative task, one model output could plausibly be interpreted as multiple molecular structures, and so it is ``easier'' for the model output to be interpreted as ``correct'' or ``valid.'' We view the task of predicting graph topology and structure with explicit hydrogens and formal charges as the most rigorous evaluation of the capabilities of generative models to fit the distribution of valid molecular structures. 

\subsection{Model Evaluation}

We report three metrics measuring the validity of generated molecular topology: percent atoms stable, percent molecules stable, and percent molecules valid. An atom is defined as ``stable'' if it has valid valency. Atomic valency is defined as the sum of bond orders that an atom is participating atom. Aromatic bonds are assigned a bond order of $1.5$. A valid valency is defined as any valency that is observed in the training data for atoms of a given element and formal charge. A molecule is counted as stable if all of its constituent atoms are stable. A molecule is considered ``valid'' if it can be sanitized by rdkit \cite{noauthor_rdkit_nodate} using default sanitization settings.

Metrics regarding the validity of molecular topology fail to capture a model's ability to reproduce reasonable molecular geometries. Therefore, we also compute the Jensen-Shannon divergence of the distribution of potential energies for molecules in the training data and molecules sampled from trained models. Potential energies are obtained from the Merck Molecular Mechanics Force-Field implemented in rdkit \cite{noauthor_rdkit_nodate}. Force-field energy cannot be obtained for molecules that cannot be sanitized by rdkit, and so the reported Jensen-Shannon divergences are for valid molecules only.

Molecule quality metrics are reported for samples of $10,000$ molecules, repeated 5 times. We report inference time for FlowMol and baseline models. We measure inference time as the time required to generate one batch of $100$ molecules on the same NVIDIA GeForce RTX 2060 GPU. This inference procedure is also repeated five times. Inference is run on FlowMol using Euler integration with $100$ evenly-spaced timesteps. All results are reported with $95\%$ confidence intervals. For all samplings, the number of atom in each molecule is sampled from the distribution of atoms in molecules from the training data.
 
\subsection{Model Ablations}

We train multiple versions of our model to evaluate the effects of several aforementioned design choices. To observe the effect of endpoint reparameterization (sec \ref{sec:ep}), we train equivalent models with both the vector-field objective \eqref{eq:cfm-loss} and the endpoint objective \eqref{eq:endpoint-objective}. We train models using SimplexFlow with all three categorical priors proposed in Section \ref{sec:priors} which have support on the simplex. To determine whether SimplexFlow improves performance, we also train models where the prior distribution for categorical features is a standard Gaussian distribution. In this setting, the generated flows are not constrained to the simplex, and it can be said that the flows do not ``respect'' the categorical nature of the data. This is similar to the atom type flows in \citet{song_equivariant_2023} and atom type diffusion in \citet{hoogeboom_equivariant_2022}. 

All of the mentioned model ablations are tested on the QM9 dataset and the results are presented in Section \ref{sec:ablation-results}. A subset of these ablations were also performed on the GEOM dataset. GEOM ablations are available in Appendix \ref{ap:geom-ablations}. None of the effects observed in GEOM ablations contradict those seen for QM9 ablations. For metrics reported in ablations, results are averaged over two identical models trained with different random seeds

\subsection{Comparison to Dirichlet Flows}
We compare SimplexFlow to concurrent work that developed Dirichlet Flows \cite{stark_dirichlet_2024} for flow matching on the simplex. Briefly, for a $d$-categorical variable $x$ represented as a point on the simplex, the conditional probability path is 

\begin{equation}
    p_t(x|x_1=e_i) = \mathrm{Dir}(x|\gamma = 1 + e_i\omega)
\end{equation}

Where $\mathrm{Dir}$ is a Dirichlet distribution parameterized by $\gamma$ and $\omega$ represents time. The Dirichlet conditional flow must start at $\omega = 1$ and only converges to $\delta(x-e_i)$ in the limit $\omega \to \infty$. In order to incorporate Dirichlet flows into our model, we define the relation $\omega_t = \omega_{\mathrm{max}}\alpha_t + 1$, where $\alpha_t$ is defined by \eqref{eq:cos-sched}. Dirichlet flow matching necessitates the use of a uniform prior over the simplex for categorical variables and so we do not experiment with other simplex priors described in Section \ref{sec:priors}.

\subsection{Baselines}

We compare FlowMol to three baselines: MiDi \cite{vignac_midi_2023}, JODO \cite{huang_learning_2023}, and EquiFM \cite{song_equivariant_2023}. MiDi and JODO perform the same generation task: predicting atom positions, atom types, formal charges, and bond orders. The key difference from FlowMol is that MiDi and JODO are diffusion models. EquiFM as described in Section \ref{sec:fmdenovo} is a flow matching model for de novo molecule generation; however, the model does not predict bond orders or atomic charges. We do not report the performance of EquiFM on the GEOM dataset because the authors have not released a model checkpoint.  

\section{Results}

\subsection{Model Ablations} \label{sec:ablation-results}

\begin{table}
    \centering
    \caption{FlowMol Ablations on QM9 with explicit hydrogens}
    \label{tab:qm9 ablation}
    
    \begin{tabular}{ll|cccc}
    \toprule
     \textbf{Flow Type} & \textbf{Categorical Prior} & \thead{\textbf{Atoms Stable} \\ \textbf{(\%) ($\uparrow$)}} & \thead{\textbf{Mols Stable} \\ \textbf{(\%) ($\uparrow$)}} & \thead{\textbf{Mols Valid} \\ \textbf{(\%) ($\uparrow$)}} & \thead{\textbf{JS(E)} \\ \textbf{($\downarrow$)}} \\
    \midrule
    Dirichlet & uniform-simplex & $ 98.4 { \scriptstyle \pm 0.0 }  $ & $ 80.0 { \scriptstyle \pm 0.3 }  $ & $ 85.5 { \scriptstyle \pm 0.3 }  $ & $ 0.15 { \scriptstyle \pm 0.01 }  $ \\
    \cline{1-6}
    \multirow[t]{4}{*}{endpoint} & uniform-simplex & $ 98.9 { \scriptstyle \pm 0.1 }  $ & $ 84.2 { \scriptstyle \pm 0.9 }  $ & $ 88.9 { \scriptstyle \pm 0.6 }  $ & $ 0.11 { \scriptstyle \pm 0.01 }  $ \\
    & marginal-simplex & $ 99.5 { \scriptstyle \pm 0.1 }  $ & $ 91.9 { \scriptstyle \pm 0.7 }  $ & $ 96.1 { \scriptstyle \pm 0.2 }  $ & $ 0.06 { \scriptstyle \pm 0.00 }  $ \\
    & barycenter & $ 99.5 { \scriptstyle \pm 0.0 }  $ & $ 91.4 { \scriptstyle \pm 0.5 }  $ & $ 93.6 { \scriptstyle \pm 0.5 }  $ & $ \mathbf{ 0.05 { \scriptstyle \pm 0.00 } } $ \\
    & Gaussian & $ \mathbf{99.7 { \scriptstyle \pm 0.0 } } $ & $ \mathbf{ 96.0 { \scriptstyle \pm 0.1 } } $ & $ \mathbf{ 96.9 { \scriptstyle \pm 0.1 } } $ & $ 0.09 { \scriptstyle \pm 0.01 }  $ \\
    \cline{1-6}
    \multirow[t]{2}{*}{vector-field} & marginal-simplex & $ 98.6 { \scriptstyle \pm 0.0 }  $ & $ 79.4 { \scriptstyle \pm 0.3 }  $ & $ 86.2 { \scriptstyle \pm 0.3 }  $ & $ 0.07 { \scriptstyle \pm 0.00 }  $ \\
    & Gaussian & $ 99.5 { \scriptstyle \pm 0.0 }  $ & $ 93.6 { \scriptstyle \pm 0.7 }  $ & $ 94.7 { \scriptstyle \pm 0.7 }  $ & $ 0.08 { \scriptstyle \pm 0.01 }  $ \\
    \cline{1-6}
    \bottomrule
    \end{tabular}

\end{table}

Results of model ablation experiments on the QM9 dataset are shown in Table \ref{tab:qm9 ablation}. Most notably, models that use SimplexFlow for categorical variables (those with categorical priors constrained to the simplex) consistently underperform models with Gaussian categorical priors. The best performing SimplexFlow model (endpoint parameterization, marginal-simplex prior) achieves $96.1\%$ valid molecules while an equivalent model using a Gaussian prior achieves $96.9\%$ valid molecules. 

Models trained under the endpoint objective achieve superior performance to otherwise identical models trained under the vector-field objective. For example, Table \ref{tab:qm9 ablation} shows that a model trained with the a marginal-simplex categorical prior obtains $79\%$ stable molecules under the vector-field objective and $92\%$ stable molecules under the endpoint objective. This is effect is also observed with models using a Gaussian categorical prior but to a lesser extent.

We find that models using Dirichlet conditional probability paths \cite{stark_dirichlet_2024} yields approximately equivalent performance to the conditional probability path \eqref{eq:flex-interpolant} with a uniform-simplex categorical prior. Among models satisfying the constraints of SimplexFlow (sec. \ref{sec:simpflow}), the uniform-simplex prior yielded the worst performance. The marginal-simplex and barycenter priors yield approximately equivalent performance. Although the models using marginal-simplex and barycenter priors produce relatively fewer valid molecules, the molecules generated by these models exhibit the lowest Jensen-Shannon divergence to the energy distribution of the training data. 

\subsection{Comparison with Baselines}

\begin{table}
    \centering
    \caption{Comparison of FlowMol to baseline models on the QM9 and GEOM-Drugs datasets}
    \label{tab:my_label}
    \begin{tabular}{l|c|ccccc}
         \toprule
         \textbf{Model} & \textbf{Dataset} & \textbf{\thead{Atoms Stable \\ (\%) ($\uparrow$)}} & \textbf{\thead{Mols Stable \\ (\%) ($\uparrow$)}} & \textbf{\thead{Mols Valid \\ (\%) ($\uparrow$) }} & \textbf{ \thead{JS(E) \\ ($\downarrow$) } } & \textbf{ \thead{Inference Time \\ (s) ($\downarrow$)}} \\
         \midrule
         JODO \cite{huang_learning_2023} & \multirow{4}{*}{QM9} & $ 99.9 { \scriptstyle \pm 0.0 }  $ & $ 98.7 { \scriptstyle \pm 0.2 }  $ & $ 98.9 { \scriptstyle \pm 0.2 }  $ & $ 0.12 { \scriptstyle \pm 0.01 }  $ & $ 116 { \scriptstyle \pm 2 }  $ \\
         MiDi \cite{vignac_midi_2023} & & $ 99.8 { \scriptstyle \pm 0.0 }  $ & $ 97.5 { \scriptstyle \pm 0.1 }  $ & $ 98.0 { \scriptstyle \pm 0.2 }  $ & $ 0.05 { \scriptstyle \pm 0.00 }  $ & $ 89 { \scriptstyle \pm 7 }  $ \\
         EquiFM \cite{song_equivariant_2023} & & $ 99.4 { \scriptstyle \pm 0.0 }  $ & $ 93.2 { \scriptstyle \pm 0.3 }  $ & $ 94.4 { \scriptstyle \pm 0.2 }  $ & $ 0.08 { \scriptstyle \pm 0.00 }  $ & $ 25 { \scriptstyle \pm 3 }  $ \\
         FlowMol (ours) & & $ 99.7 { \scriptstyle \pm 0.0 }  $ & $ 96.2 { \scriptstyle \pm 0.1 }  $ & $ 97.3 { \scriptstyle \pm 0.1 }  $ & $ 0.08 { \scriptstyle \pm 0.00 }  $ & $ 6 { \scriptstyle \pm 0 }  $ \\
         \midrule
         JODO \cite{huang_learning_2023} & \multirow{3}{*}{GEOM-Drugs} & $ 99.8 { \scriptstyle \pm 0.0 }  $ & $ 90.7 { \scriptstyle \pm 0.5 }  $ & $ 76.5 { \scriptstyle \pm 0.8 }  $ & $ 0.17 { \scriptstyle \pm 0.01 }  $ & $ 235 { \scriptstyle \pm 16 }  $ \\
         MiDi \cite{vignac_midi_2023} & & $ 99.0 { \scriptstyle \pm 0.2 }  $ & $ 85.1 { \scriptstyle \pm 0.9 }  $ & $ 71.6 { \scriptstyle \pm 0.9 }  $ & $ 0.23 { \scriptstyle \pm 0.00 }  $ & $ 754 { \scriptstyle \pm 119 }  $ \\
         FlowMol (ours) & & $ 99.0 { \scriptstyle \pm 0.0 }  $ & $ 67.5 { \scriptstyle \pm 0.2 }  $ & $ 51.2 { \scriptstyle \pm 0.3 }  $ & $ 0.33 { \scriptstyle \pm 0.01 }  $ & $ 22 { \scriptstyle \pm 1 }  $ \\
         \bottomrule
    \end{tabular}

\end{table}

FlowMol achieves superior performance to EquiFM \cite{song_equivariant_2023} on QM9; for example, it produces $3\%$ more valid molecules while having equivalent divergence to the training data energy distribution. FlowMol approaches the performance of diffusion baselines (JODO, MiDi) on QM9 but does not perform as well on the GEOM-Drugs dataset. The fact that fewer generated molecules are valid on the GEOM-Drugs dataset cannot be attributed solely to the difference in molecule sizes between the two datasets, because FlowMol's atom-level stability is also worse for GEOM-Drugs than QM9 (99.0\% on GEOM vs 99.7\% on QM9). Despite the fact that MiDi and FlowMol achieve equivalent atom-level stability ($99.0\%$), MiDi produces significantly more topologically correct molecules. For example, FlowMol achieves $68\%$ stable molecules while MiDi achieves $85\%$. 

FlowMol exhibits substantially faster inferences times than all baseline models. This difference is primarily due to the fewer number of integration steps needed by FlowMol. We find empirically that sample quality does not improve when using more than 100 integration steps. JODO, MiDi, and EquiFM use 1000 integration steps by default. The need for fewer integration steps than diffusion models is a recognized advantage of flow matching models over diffusion \cite{liu_flow_2022,tong_improving_2023}.

\section{Discussion}

FlowMol improves upon the existing state of the art flow matching method for molecule generation; however, it still does not outperform diffusion models trained for the same task. A key difference between FlowMol and the diffusion baselines presented here is that the conditional trajectories are deterministic in FlowMol and stochastic in diffusion models. Prior works have presented theoretical \cite{albergo_stochastic_2023} and empirical \cite{stark_harmonic_2024} evidence that stochastic conditional trajectories yield improved model performance. 

Our results raise interesting questions about the design of prior distributions for flow matching models. Our intuition was that a stronger prior that is ``closer'' to the data distribution would yield more faithful recapitulation of the target distribution. The results of our model ablations suggest this intuition is incorrect. The next natural questions are: why is a Gaussian prior the most performant of those tested here? and what are the qualities of a prior that best enable recapitulation of the target distribution? A possible explanation for our results is a dependence on the ``volume'' of the prior. Empirically when the prior for categorical features has support on a small number of unique values, the model fails to produce any valid molecules. Adding a ``blur'' as described in Section \ref{sec:priors} dramatically improves model performance. Correspondingly, priors constrained to the simplex reliably yield poorer performance than Gaussian priors; these observations could all be explained through the perspective of the prior's capacity for serving as one domain of a homeomorphism to a more complex distribution.

Another explanation for the superiority of Gaussian priors may involve the shape of conditional trajectories induced by the prior. Conditional trajectories are more likely to intersect when constrained to a smaller space, such as the simplex. This explanation is also supported by the observation that the marginal-simplex and barycenter priors yield substantially improved performance over uniform-simplex priors. \citet{tong_improving_2023} suggest that sampling conditional pairs $(g_0,g_1)$ from an optimal transport (OT) alignment $\pi(g_0,g_1)$ improves performance precisely because the marginal vector field yields straighter lines with fewer intersections. In this work, an OT plan is computed but only for atomic positions. Perhaps computing an OT alignment over the product space of all the data modalities represented here could alleviate this issue.

\section{Conclusions}

FlowMol is the first generative model to jointly sample the topological and geometric structure of small molecules. FlowMol improves upon existing flow matching models for molecule generation and achieves competitive performance with diffusion-based models while exhibiting inference speeds an order of magnitude faster. We present a method for flow matching on categorical variables, SimplexFlow, and demonstrate that constraining flows to a smaller space does not yield performance benefits. We think this result raises interesting and relevant questions about the design of flow matching for mixed continuous/categorical generative tasks and provide potential hypotheses to begin exploring in future work.

\section{Acknowledgements}

We thank Rishal Aggarwal, Gabriella Gerlach, and Daniel Peñahererra for useful feedback and discussions.

This work is funded through R35GM140753 from the National Institute of General Medical Sciences. The content is solely the responsibility of the authors and does not necessarily represent the official views of the National Institute of General Medical Sciences or the National Institutes of Health.

\typeout{} 
\bibliography{simplexflow}

\appendix

\section{Proof that the simplex is closed under linear interpolation} \label{ap:closed simplex}

Assume we have access to two vectors on the simplex $a,b \in \mathcal{S}^d$. We define another point $c$ by linear interpolation between $a$ and $b$:

\begin{equation}
    c := ta + (1-t)b  \quad \textrm{  where } t \in [0,1]
\end{equation}

Therefore, each entry of $c$ can be written as $c_i = ta_i + (1-t)b_i$. Given that $a_i, b_i \geq 0$ by definition, $c_i$ must also be positive. We can also write the the sum of the entries of $c$ as:

\begin{align}
    \sum_{i=1}^d c_i &= \sum_{i=1}^d ta_i + (1-t)b_i \\
    &= t \sum_{i=1}^d a_i + (1-t)\sum_{i=1}^d b_i \\
    &= t + (1-t) = 1
\end{align}

We have shown that $c_i \geq 0$ and $\sum_{i=1}^d c_i = 1$. Therefore $c \in \mathcal{S}^d$ and we can conclude that $\mathcal{S}^d$ is closed under linear interpolation.

\section{Proof that flows are on the simplex} \label{ap:simplex-proof} 

We will define a flow matching model for a categorical variable $y$ that can take one of $d$ discrete values. The target distribution for $y$, $p_1(y)$ is a mixture of point masses on the vertices of $\mathcal{S}^d$. We add the constraint that $p_0(y)$ only has support on $\mathcal{S}^d$.


We choose conditional trajectories of the form \eqref{eq:flex-interpolant} and a cosine interpolant schedule of the form \eqref{eq:cos-sched}. We train a neural network to minimize the endpoint objective \eqref{eq:endpoint-objective}, and as a result we have access to a model which predicts the endpoint of our trajectory given a current position: $\hat{y}_1(y_t)$.

To generate samples from $p_1(y)$, we first sample $y_0 \sim p_0(y)$, and integrate the ODE

\begin{equation} \label{eq:catode}
    \frac{dy}{dt} = u_\theta(y) = \frac{\alpha'_t}{1-\alpha_t}(\hat{y}_1 - y_t) 
\end{equation}

Integration via Euler's method would produce trajectories according to:

\begin{equation} \label{eq:euler}
    y_s = y_t + u_\theta(y_t)(s-t)
\end{equation}

Where $s = t + \Delta t$. In order to prove that all trajectories produced by our vector field lie on the simplex, it would be sufficient to prove that $y_s \in \mathcal{S}^d$ in the limit as $\Delta t \to 0$.  

Substituting \eqref{eq:catode} into \eqref{eq:euler} yields the following update rule for integration by Euler's method:

\begin{equation} \label{eq:flex-update-rule}
    y_{s} = \frac{\alpha'_t(s-t)}{1-\alpha_t}\hat{y}_1 + \frac{1-\alpha_t - \alpha'_t(s-t)}{1-\alpha_t}y_t
\end{equation}

Relying on the property that the simplex is closed under linear interpolation (Appendix \ref{ap:closed simplex}), our strategy is to prove that integrating trajectories via  \eqref{eq:flex-update-rule} results in recursive applications of linear interpolation between two points on the simplex.

More formally, the right hand side of \eqref{eq:flex-update-rule} would be linear interpolation between two points on the simplex if the following conditions were satisfied:

\begin{enumerate}
    \item $\frac{\alpha'_t(s-t)}{1-\alpha_t} + \frac{1-\alpha_t - \alpha'_t(s-t)}{1-\alpha_t} = 1 $
    \item $\frac{\alpha'_t(s-t)}{1-\alpha_t} \in [0,1]$
    \item $\hat{y}_1(y_t), y_t  \in \mathcal{S}^d$
\end{enumerate}

The first condition is obviously true.

The second condition can be written as two inequalities $0 \leq \frac{\alpha'_t(s-t)}{1-\alpha_t} \leq 1$. The first inequality reduces to $\alpha'_t \geq 0 $; the interpolant must be monotonically increasing. The second inequality can be seen as an upper bound on the step size that can be used during integration:

\begin{equation} \label{eq:intub}
    s-t \leq \frac{1 - \alpha_t}{\alpha'_t}
\end{equation}

For well-behaved interpolant schedules and many reasonable choices of interpolant, this inequality is satisfied in the limit as $s-t \to 0$. 

Regarding the third condition: by definition, $\hat{y}_1 \in \mathcal{S}^d$; this is practically enforced by placing softmax activations on the output of the neural network. If $y_0 \in \mathcal{S}^d$, a condition which is guaranteed by our choice of prior $p_0(y)$, then the first application of the update rule \eqref{eq:flex-update-rule} would satisfy all three conditions and as a result $y_{0+\Delta t} \in \mathcal{S}^d$. By induction, every subsequent application of the update rule \eqref{eq:flex-update-rule} would yield an integration step that is linear interpolation between two points on the simplex. 

As a result, all trajectories generated by the ODE \label{eq:cat-ode} will lie on the simplex in the limit of a infinitely small integration step. And, in practice, infinitely small integration steps are not actually necessary to yield trajectories on the simplex. More on this in Appendix \ref{ap:practical inteprolants}.

\section{Intepolation Schedules and Integration Step Sizes} \label{ap:practical inteprolants}

\begin{figure}[h]
    \centering
    \includegraphics[width=0.75\textwidth]{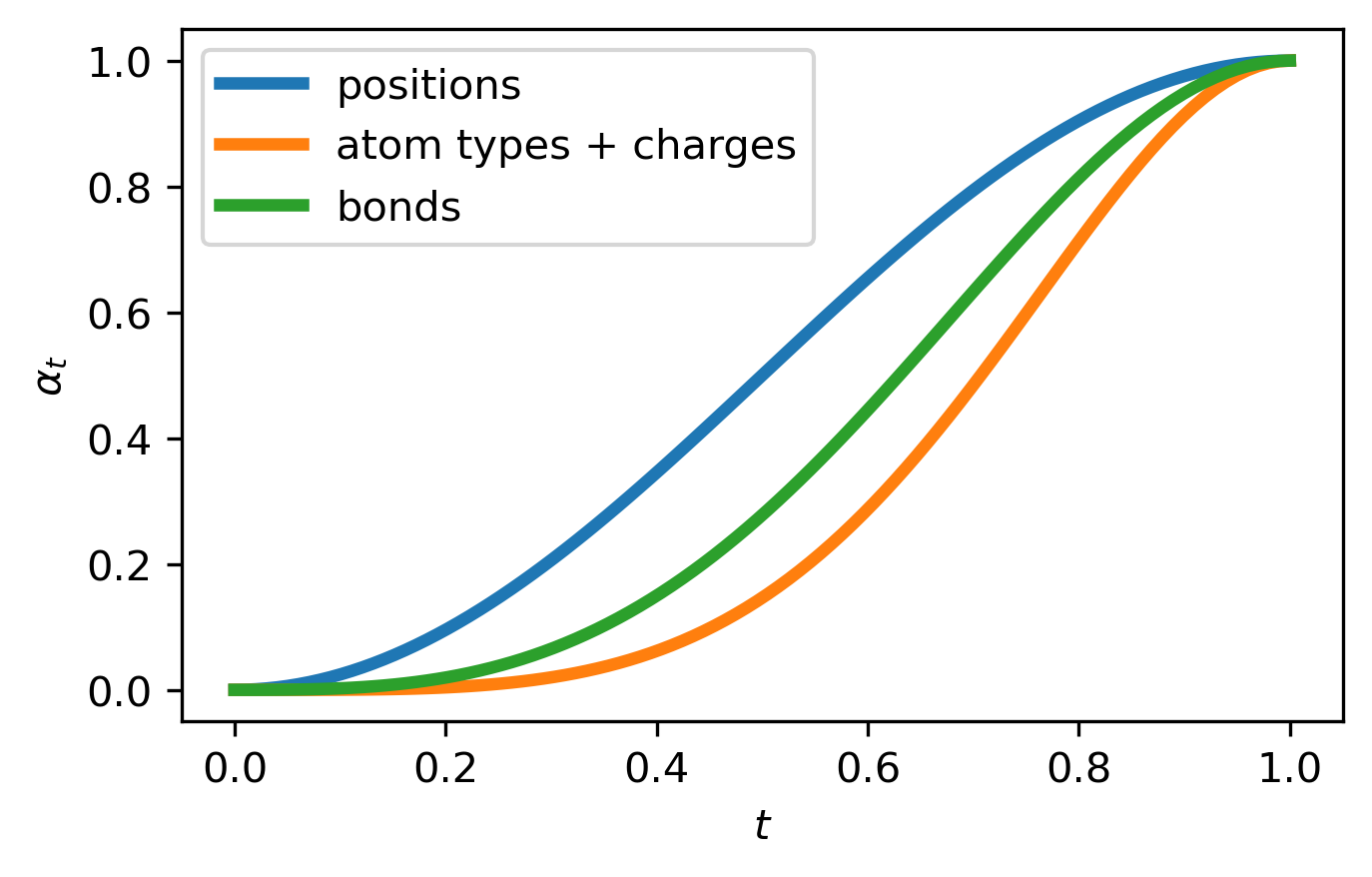}
    \caption{Interpolant Schedules for the QM9 Dataset}
    \label{fig:interp-sched}
\end{figure}

The relative rates at which molecular features are generated are determined by setting values of the parameter $\nu$ in the cosine interpolant schedule \eqref{eq:cos-sched}. For the QM9 dataset we set $\nu = (\nu_X, \nu_A, \nu_C, \nu_E) = (1, 2, 2, 1.5)$. For the GEOM-Drugs dataset this is set to $(1, 2, 2, 2)$. These are the same values used for the cosine noise schedule in \citet{vignac_midi_2023}. The interpolant schedules used for the QM9 dataset are plotted in Figure \ref{fig:interp-sched}.

\begin{figure}[h]
    \centering
        \includegraphics[width=0.45\textwidth]{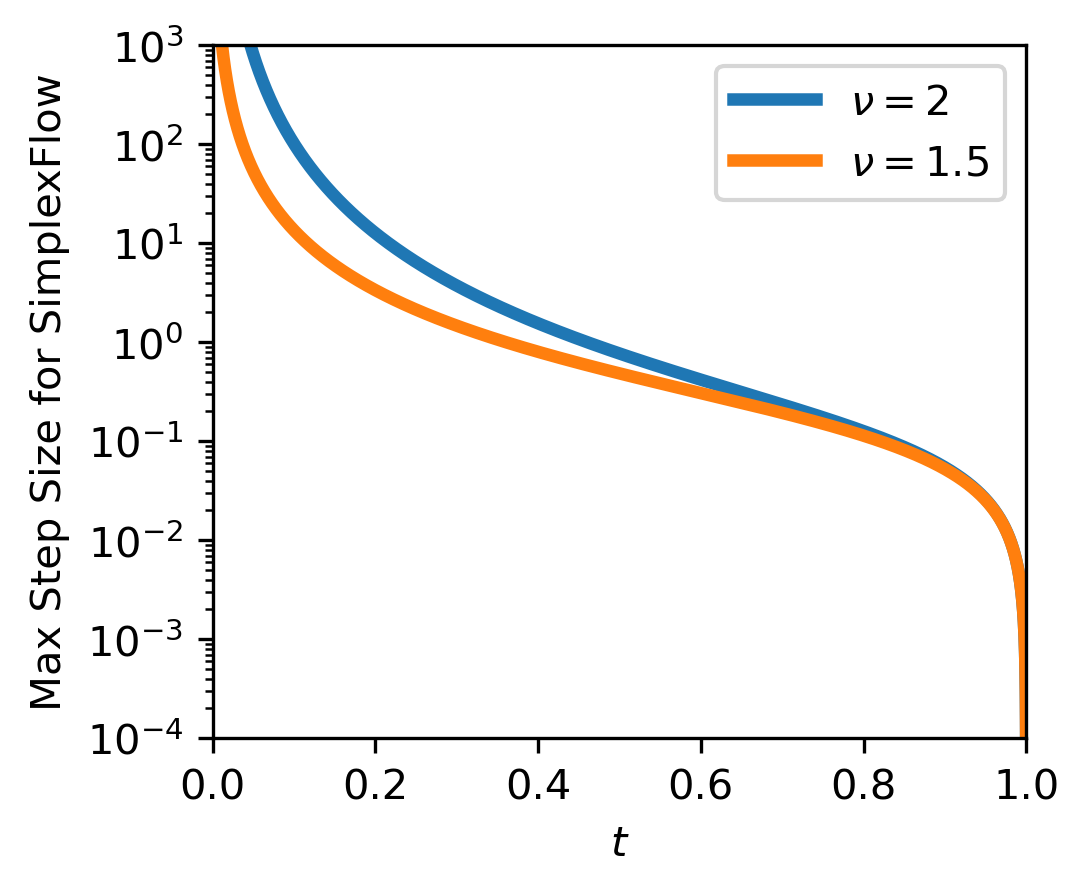}
        \includegraphics[width=0.45\textwidth]{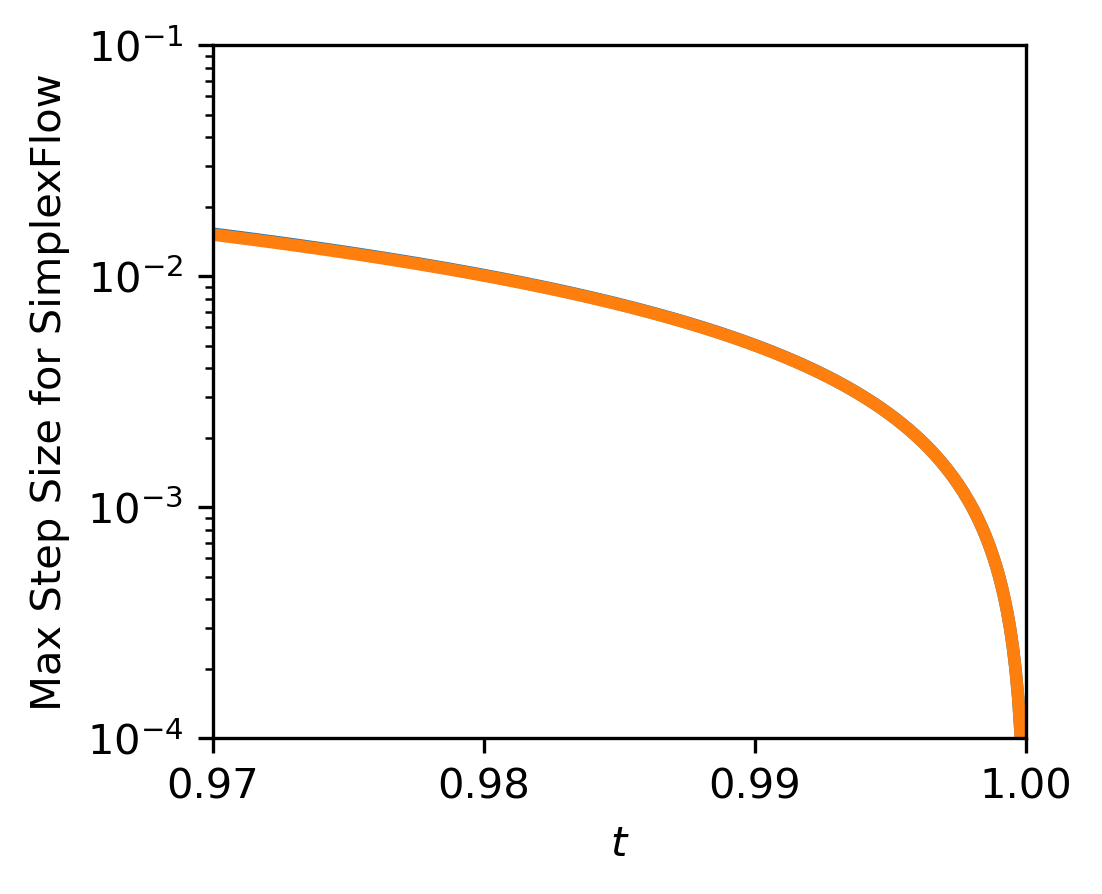}
    \caption{\textbf{Left:} maximum integration step size to remain on the simplex for a cosine interpolant. \textbf{Right:} zoomed in view of the asymptotic decline of the maximum step size as $t \to 1$}
    \label{fig:stepsize}
\end{figure}

In Appendix \ref{ap:simplex-proof} we derive an upper bound on the integration step size that can be used that guarantees SimplexFlow trajectories will remain on the simplex \eqref{eq:intub}. In Figure \ref{fig:stepsize} we plot this maximum step size as a function of $t$ for for cosine interpolation schedules \eqref{eq:cos-sched}. For the results presented in this paper we sample molecules by performing Euler intergration with 100 evenly-spaced integration steps. This corresponds to a constant step size of $10^{-2}$. According to Figure \ref{fig:stepsize}, this step size ensures trajectories will remain on the simplex until approximately $t=0.98$.
 
\section{Model Architecture} \label{ap:arch}

FlowMol is implemented using PyTorch and the Deep Graph Library (DGL) \cite{wang_deep_2020}.

Each node is endowed with a position in space $x_i \in \mathbb{R}^3$, scalar features $s_i \in \mathbb{R}^d$, and vector features $v_i \in \mathbb{R}^{c \times 3}$. Scalar features are initialized at the network input by concatenating atom type and charge vectors: $s_i^{(0)} = [a_i:c_i]$. Vector features are initialized to zeros $v_i^{(0)} = \mathbf{0}$.
Each edge is endowed with scalar edge features that, at the input to the network, are the bond order at time $t$. We enforce that the bond order on both edges for a pair of atoms is identical: $e_{ij}^{(0)} = e_{ji}^{(0)}$.

\paragraph{Molecule Update Block} We define a Molecule Update Block which will update all graph features $x_i, s_i, v_i, e_{ij}$. Each molecule update block is comprised of 3-sub blocks: a node feature update block, a node position update block, and an edge feature update block. The input molecule graph is passed through $L$ Molecule Update Blocks. Vector features are operated on by geometric vector perceptions (GVPs). A detailed description of our implementation of GVP is provided in Section \ref{sec:gvp}.

\paragraph{Node Feature Update Block} The node feature update block will perform a graph convolution to update node scalar and vector features $s_i, v_i$. The message generating and node-update functions for this graph convolution are each chains of GVPs. GVPs accept and return a tuple of scalar and vector features. Therefore, scalar and vector messages $m_{i \to j}^{(s)}$ and $m_{i \to j}^{(v)}$ are generated by a single function $\psi_M$ which is two GVPs chained together.

\begin{equation}
    \label{gvp_mij}
    m_{i \to j}^{(s)}, m_{i \to j}^{(v)} = \psi_M \left(
    \left[ s_i^{(l)} : e_{ij}^{(l)} : d_{ij}^{(l)} \right]
    , \left[v_i : \frac{x_i^{(l)} - x_j^{(l)}}{d_{ij}^{(l)}} \right]\right) 
\end{equation}

Where $:$ denotes concatenation, and $d_{ij}$ is the distance between nodes $i$ and $j$ at molecule update block $l$. In practice, we replace all instances of $d_{ij}$ with a radial basis embedding of that distance before passing through GVPs or MLPs. Message aggregation and node features updates are performed as described in \cite{jing_equivariant_2021}:

\begin{equation}
    \label{gvp_update}
    s_i^{(l+1)}, v_i^{(l+1)} = \mathrm{LayerNorm} \left( [s_i^{(l)}, v_i^{(l)}] + \\
     \psi_U \left( \frac{1}{|\mathcal{N}(i)|} \sum_{j \in \mathcal{N}(i) } \left[ m_{j \to i}^{(s)}, m_{j \to i}^{(v)} \right]   \right) \right) 
\end{equation}

The node update function $\psi_U$ is a chain of three GVPs.

\paragraph{Node Position Update Block} The purpose of this block is to update node positions $x_i$. Node positions are updated as follows:

\begin{equation}
    x_i^{(l+1)} = x_i^{(l)} + \psi_x \left( s_i^{(l+1)}, v_i^{(l+1)} \right)
\end{equation}

Where $P$ is a chain of 3 GVPs in which the final GVP emits 1 vector and 0 scalar features. Moreover for the final GVP, the vector-gating activation function ($\sigma_g$ in Algorithm \ref{alg:gvp}), which is typically a sigmoid function, is replaced with the identity.

\paragraph{Edge Feature Update Block} Edge features are updated by the following equation:

\begin{equation}
       e_{ij}^{(l+1)} = \mathrm{LayerNorm} \left(e_{ij}^{(l)} + \phi_e \left( s_i^{(l+1)}, s_j^{(l+1)}, d_{ij}^{(l+1)} \right) \right)
\end{equation}

Where $\phi_e$ is a shallow MLP that accepts as input the node scalar features of nodes participating in the edge as well as the distance between the nodes from the positions compute in the NPU block.

\subsection{GVP with Cross Product} \label{sec:gvp}

A geometric vector perception (GVP) can be thought of as a single-layer neural network that applies linear and point-wise non-linear transformation to its inputs. The difference between GVP and a conventional feed-forward neural network is that GVPs operate on two distinct data types: scalars and vectors. GVP also allows these data types to exchange information while preserving equivariance of the output vectors. The original GVP only applied linear transformations to the vector features and as a result produces output vectors that are E(3)-equivariant. 

We introduce a modification to the GVP as its presented in \citet{jing_equivariant_2021}; specifically we perform a cross product operation on the input vectors. The motivation for this is that the cross product is \textit{not} equivariant to reflections. As a result, the version of GVP we present here is SE(3) equivariant. The operations for our cross product enhanced GVP are described in Algorithm \ref{alg:gvp}.

\begin{algorithm}
\caption{Geometric Vector Perceptron with Cross Product} 
\label{alg:gvp}
\begin{algorithmic}
\setstretch{1.2}
\STATE
\STATE \textbf{Input:} Scalar and vector features: $(s,v) \in \mathbb{R}^f \times \mathbb{R}^{\nu \times 3}$
\STATE \textbf{Output:} Scalar and vector features: $(s',v') \in \mathbb{R}^j \times \mathbb{R}^{\mu \times 3}$
\STATE \textbf{Hyperparameter:} Number of hidden vector features $n_h \in \mathbb{Z}^{+}$ 
\STATE \textbf{Hyperparameter:} Number of cross product features $n_{cp} \in \mathbb{Z}^{+}$ 
\STATE \ \ \ \ $v_{h} \gets W_h v \quad \in \mathbb{R}^{n_h \times 3} $
\STATE \ \ \ \ $v_{cp} \gets W_{cp} v \quad \in\mathbb{R}^{2n_{cp} \times 3}$
\STATE \ \ \ \ $v_{cp} \gets v_{cp}[:n_{cp}] \times v_{cp}[n_{cp}:] \quad \in \mathbb{R}^{n_{cp} \times 3}  $ // cross product
\STATE \ \ \ \ $ v_{h+cp} \gets \mathrm{Concat}(v_h, v_{cp}) \quad \in \mathbb{R}^{(n_h + n_{cp}) \times 3} $ // concatenation along rows
\STATE \ \ \ \ $v_{\mu} \gets W_{\mu} v_{h+cp} \quad \in \mathbb{R}^{\mu \times 3} $
\STATE \ \ \ \ $ s_{h+cp} \gets \norm{v_{h+cp}} \quad \in\mathbb{R}^{n_h + n_{cp}} $ 
\STATE \ \ \ \ $ s_{f+h+cp} \gets \mathrm{Concat}(s, s_{h+cp})$'
\STATE \ \ \ \ $s_j \gets W_j s_{f+h+cp} + b_j \quad \in \mathbb{R}^j$
\STATE \ \ \ \ $s' \gets \sigma(s_j) \quad \in \mathbb{R}^{j}$
\STATE \ \ \ \ $v' \gets { \sigma_g\left(W_g[\sigma^+(s_m)]+b_g\right)\odot v_\mu} \ (\text{row-wise}) \quad \in \mathbb{R}^{\mu \times 3}$ 
\STATE  \textbf{return} $(s', v')$  
\end{algorithmic}
\end{algorithm}

\section{Model Ablations on GEOM Dataset} \label{ap:geom-ablations}

\begin{table}[H]
    \centering
    \caption{FlowMol ablations on GEOM-Drugs with explicit hydrogens}
    \label{tab:geom}
    
    \begin{tabular}{ll|cccc}
    \toprule
     \textbf{Flow Type} & \textbf{Categorical Prior} & \thead{\textbf{Atoms Stable} \\ \textbf{(\%) ($\uparrow$)}} & \thead{\textbf{Mols Stable} \\ \textbf{(\%) ($\uparrow$)}} & \thead{\textbf{Mols Valid} \\ \textbf{(\%) ($\uparrow$)}} & \thead{\textbf{JS(E)} \\ \textbf{($\downarrow$)}} \\
    \midrule
    Dirichlet & uniform-simplex & $ 94.6 { \scriptstyle \pm 0.2 }  $ & $ 18.7 { \scriptstyle \pm 0.2 }  $ & $ 14.0 { \scriptstyle \pm 1.4 }  $ & $ 0.46 { \scriptstyle \pm 0.01 }  $ \\
    \cline{1-6}
    \multirow[t]{3}{*}{endpoint} & marginal-simplex & $ 97.7 { \scriptstyle \pm 0.0 }  $ & $ 36.3 { \scriptstyle \pm 0.1 }  $ & $ 28.3 { \scriptstyle \pm 0.3 }  $ & $ 0.36 { \scriptstyle \pm 0.01 }  $ \\
    & barycenter & $ 97.6 { \scriptstyle \pm 0.0 }  $ & $ 35.2 { \scriptstyle \pm 0.4 }  $ & $ 27.6 { \scriptstyle \pm 0.5 }  $ & $ 0.30 { \scriptstyle \pm 0.01 }  $ \\
    & Gaussian & $ 98.9 { \scriptstyle \pm 0.0 }  $ & $ 66.3 { \scriptstyle \pm 1.2 }  $ & $ 49.9 { \scriptstyle \pm 1.0 }  $ & $ 0.34 { \scriptstyle \pm 0.01 }  $ \\
    \cline{1-6}
    \bottomrule
    \end{tabular}

\end{table}

\section{Training Details and Hyperparameter Choices}

QM9 models are trained with 8 Molecule Update Blocks while GEOM models are trained with 5. Atoms contain 256 hidden scalar features and 16 hidden vector features. Edges contain 128 hidden features. QM9 models are trained for 1000 epochs and GEOM models are trained for 20 epochs. QM9 models are trained on a single L40 GPU with a batch size of 64. GEOM models are trained on 4xL40 GPUs with a per-GPU batch size of 16. QM9 models train in about 3-4 days while GEOM models take 4-5 days. All model hyperparameters are visible in the config files provided in our github repository. 

\end{document}